\documentclass[twocolumn,amsmath,amssymb,superscriptaddress]{revtex4}

\usepackage[pdftex]{graphicx}

\begin{document}

\title{Fermi surface of an important nano-sized metastable phase: Al$_3$Li}

\author{J.~Laverock}
\affiliation{H.~H.~Wills Physics Laboratory, University of Bristol, Tyndall
Avenue, Bristol BS8 1TL, United Kingdom}

\author{S.B.~Dugdale}
\affiliation{H.~H.~Wills Physics Laboratory, University of Bristol, Tyndall
Avenue, Bristol BS8 1TL, United Kingdom}

\author{M.A.~Alam}
\affiliation{H.~H.~Wills Physics Laboratory, University of Bristol, Tyndall
Avenue, Bristol BS8 1TL, United Kingdom}

\author{M.V.~Roussenova}
\affiliation{H.~H.~Wills Physics Laboratory, University of Bristol, Tyndall
Avenue, Bristol BS8 1TL, United Kingdom}

\author{J.~Wensley}
\affiliation{H.~H.~Wills Physics Laboratory, University of Bristol, Tyndall
Avenue, Bristol BS8 1TL, United Kingdom}

\author{J.~ Kwiatkowska}
\affiliation{H. Niewodnicza\'{n}ski Institute of Nuclear Physics, Polish
Academy of Sciences, Radzikowskiego 152, 31-342 Krak\'{o}w, Poland}

\author{N.~Shiotani}
\affiliation{KEK-PF, Tsukuba, Ibaraki 305-0801, Japan}

\begin{abstract}
Nanoscale particles embedded in a metallic matrix are of considerable interest
as a route towards identifying and tailoring material properties.
We present a detailed investigation of the electronic structure, and in
particular the Fermi surface, of a nanoscale phase ($L1_2$ Al$_3$Li) that
has so far been inaccessible with conventional techniques, despite playing
a key role in determining the favorable material properties of the alloy
(Al\nobreakdash-9~at.~\%\nobreakdash-Li). The ordered precipitates only form
within the stabilizing Al matrix and do not exist in the bulk; here, we take
advantage of the strong positron affinity of Li to directly probe the Fermi
surface of Al$_3$Li. Through comparison with band structure calculations,
we demonstrate that the positron uniquely probes these precipitates, and
present a `tuned' Fermi surface for this elusive phase.
\end{abstract}

\maketitle

In recent years, many of the properties of the Al-Li alloy system have 
come under careful
scrutiny, not least because of the widespread interest in these materials by the
aerospace industry. In the Al-rich region of the phase diagram (Li
concentrations between 5\% and 25\%), these alloys offer high stiffness and
superior strength-to-weight ratios, principally due to the hardening which
occurs through the precipitation of nanoscale particles.  The Li-rich
strengthening precipitates, known as the $\delta'$ phase, are Al$_3$Li and are
highly ordered with an $L1_2$ structure, and remain crystallographically
coherent with the parent (face-centered cubic, fcc) solid-solution matrix with
small lattice mismatch \cite{noble1971etc,sluiter1990}.  The size of the
precipitates, and the volume they occupy, depend not only on the Li
concentration, but also the particular conditions experienced (such as heat
treatment and ageing) \cite{krug2008+radmilovic2008};
for the Al-9at.\%Li samples considered here, that volume
is approximately 20\%, comprising roughly spherical precipitates with an average
diameter of about 20nm.

However, the $\delta'$ phase is metastable \cite{sluiter1990}, and only exists
within the parent Al fcc matrix. For this reason, knowledge of the electronic
structure of the Al$_3$Li precipitates has so far only come from
band-theoretical calculations (see, for example, \cite{guo1990}).  Although some
of the strengthening qualities orginate from the fact that the precipitates act
as pinning centres for defects, it is believed that Al$_3$Li has a particularly
high Young's modulus, which, of course, stems from its electronic structure
\cite{guo1991,mikkelsen2001}.

Advantage is taken in this study of the strong positron affinity of Li-rich
regions \cite{puska1989} to directly probe the electronic
structure (in the form of its Fermi surface) of
the $\delta'$-phase Al$_3$Li precipitates. Here, the strong positron affinity 
leads to the trapping of most of the positrons in the precipitates facilitating
an unambiguous probing of the properties of the precipitates alone since the
experimental signatures are sufficiently different from the host matrix.
Previous pioneering positron studies
of nanoscale precipitates \cite{nagai2001+nagai2009,asokakumar2002} (and even
quantum dots
\cite{eijt2006}) have, for a variety of reasons, restricted their analyses to
p-space rather than k-space.  Here, the Fermi surface
(FS) of the precipitate itself is the objective.

Positron annihilation is a well-established technique for investigating the
occupution densities in k-space, and hence the FS, which is
accessed via the momentum distribution measured by the 2-Dimensional Angular
Correlation of electron--positron Annihilation Radiation (2D-ACAR) technique
\cite{west1995}.  A 2D-ACAR measurement yields a 2D projection (integration over
one dimension) of the underlying electron-positron momentum density,
$\rho^{2\gamma}({\bf p})$, in which the FS is expressed through discontinuities
in the distribution at the Fermi momenta ${\bf p}_{\rm F} = {\bf k}_{\rm F} +
{\bf G}$, where ${\bf k}_{\rm F}$ represent the loci of the FS in k-space and
${\bf G}$ is a vector of the reciprocal lattice. When the FS is of paramount
interest, the application of the Lock-Crisp-West (LCW) procedure \cite{lock1973}
is used to superimpose the contribution from successive Brillouin zones (BZ)
into the first BZ, thereby directly providing a map of the projection of the
occupied states in the BZ (i.e.\ a projection of the FS).

\begin{figure}[t!] 
\begin{center}
\includegraphics[width=1.0\linewidth]{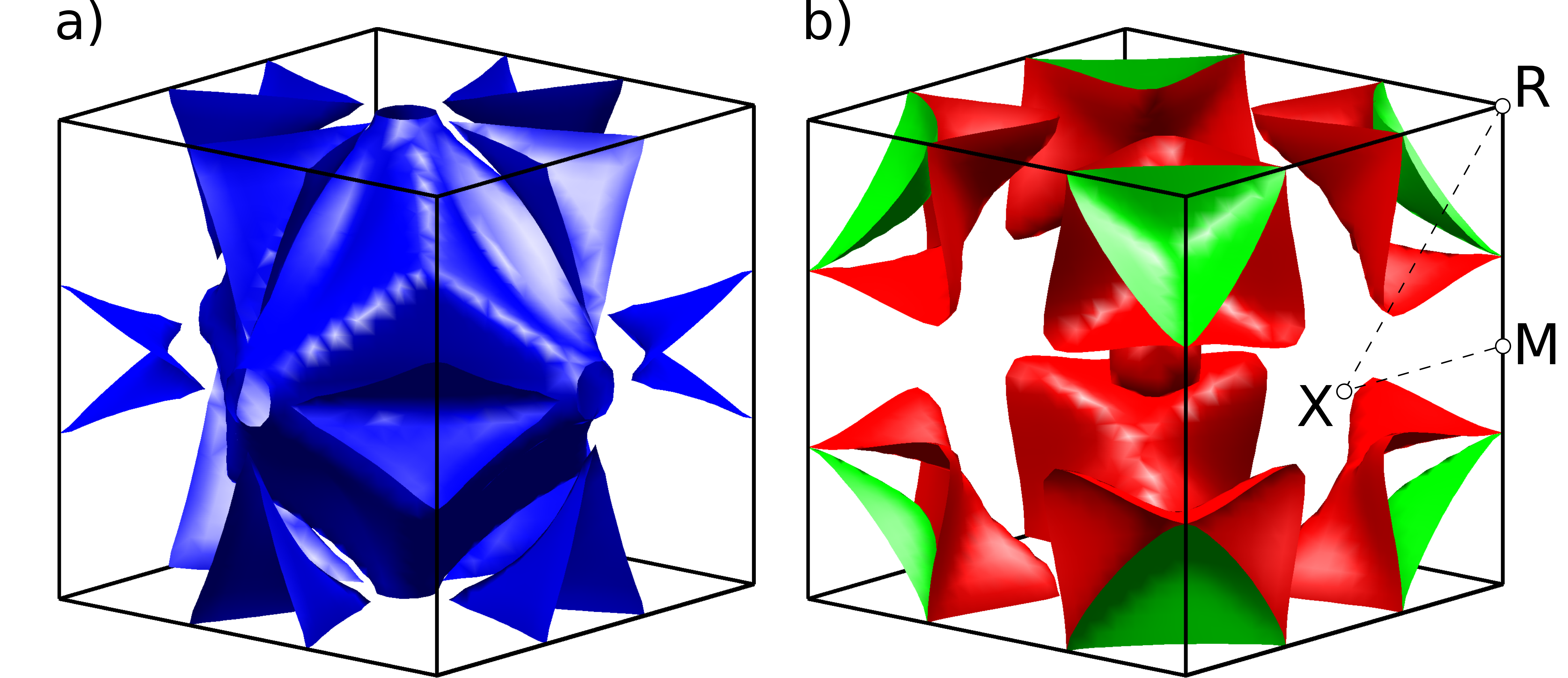}
\end{center}
\vspace*{-.25in}
\caption{The hole (a) and two electron (b) FS sheets of Al$_3$Li predicted by
the LMTO calculation. The symmetry points of the BZ are labelled in (b); the
$\Gamma$-point is at the centre.}
\label{f:fs}
\end{figure}

In order to assess the topology of the measured FS, positron annihilation
(2D-ACAR) measurements have been combined with {\em ab initio} electronic
structure calculations. The aim of this combined approach is to: (a) establish
that the measured Fermi surface indeed arise from the precipitates alone and
(b) obtain as accurate a picture as possible of the first experimental FS of
the Al$_3$Li phase. To 
achieve these goals, first of all, we consider the annihilation of the
positron, and in particular its sensitivity to the ordered precipitates, by
considering 
all possible scenarios: i) the positron annihilates {\em
only} with delocalized electrons in the $\delta'$ precipitates, and our measured
FS is that of $L1_2$ Al$_3$Li, ii) the positron annihilates with the Al matrix,
and our measured FS resembles pure Al, iii) the positron annihilates from both
the Al$_3$Li precipitates {\em and} the Al matrix, and our FS measurement is a
weighted average of both, iv) our measurements reflect the stoichiometry of the
sample, and the FS resembles that of the disordered alloy
Al$_{0.91}$Li$_{0.09}$. Electronic structure calculations of pure Al and
Al$_3$Li have been performed using the linear muffin-tin orbital (LMTO) method
\cite{lmto}, whereas for the disordered Al$_{0.91}$Li$_{0.09}$ and
Al$_{0.75}$Li$_{0.25}$ alloys, the Korringa-Kohn-Rostoker (KKR) within the
coherent potential approximation (CPA) framework
\cite{kkr} was employed. Full-potential calculations (using the Elk code
\cite{elk}) have also been performed to investigate any inaccuracies associated
with the potential shape approximations employed by the LMTO and KKR
implementations. The LMTO calculations predict three FS sheets for $L1_2$
Al$_3$Li, shown in Fig.~\ref{f:fs}, the first of which is a hole sheet
(enclosing unfilled states) and the second and third are both electron sheets
(enclosing filled electron states). Note that all FS sheets enclose filled
states (and are therefore occupied) at the
$R$-point (corner) of the BZ. These calculations agree well with previous
calculations \cite{guo1990}, as well as with our
additional KKR and Elk calculations (not presented here). Moreover, the Elk
calculations yield a large (zero-temperature) Young's modulus of 137~GPa,
in excellent agreement with the findings of Ref.~\cite{guo1991}.

A single crystal of the Al-Li alloy was grown by the Bridgman method from high 
purity Al (99.999\%) and Li (99.95\%) by
adding Li into molten Al, followed by stirring and casting, sealed within a
stainless steel capsule under an Ar
pressure of 3~atm.\ to prevent excessive loss of Li during the crystallisation
process. The final Li content in the single crystal was determined by Atomic
Absorption Spectrometry and independently in each sample by Proton Induced Gamma
Emission and was found to be 9~at.\%. The crystal was oriented using Laue
back-reflection and specimens of 1.5~mm thickness were cut along planes normal
to the main crystallographic directions. The usual procedure of alternate
mechanical
polishing and chemical polishing was applied to remove damaged surface layers.
Separate measurements on the same samples of the positron lifetime
\cite{jerzy} indicate that the positron is fully trapped in (and therefore
annihilates from) the Al$_3$Li $\delta'$ precipitates \cite{rio1994}, 
already hinting at scenario (i) outlined above.

\begin{figure}[t!] 
\begin{center}
\includegraphics[width=1.0\linewidth,clip]{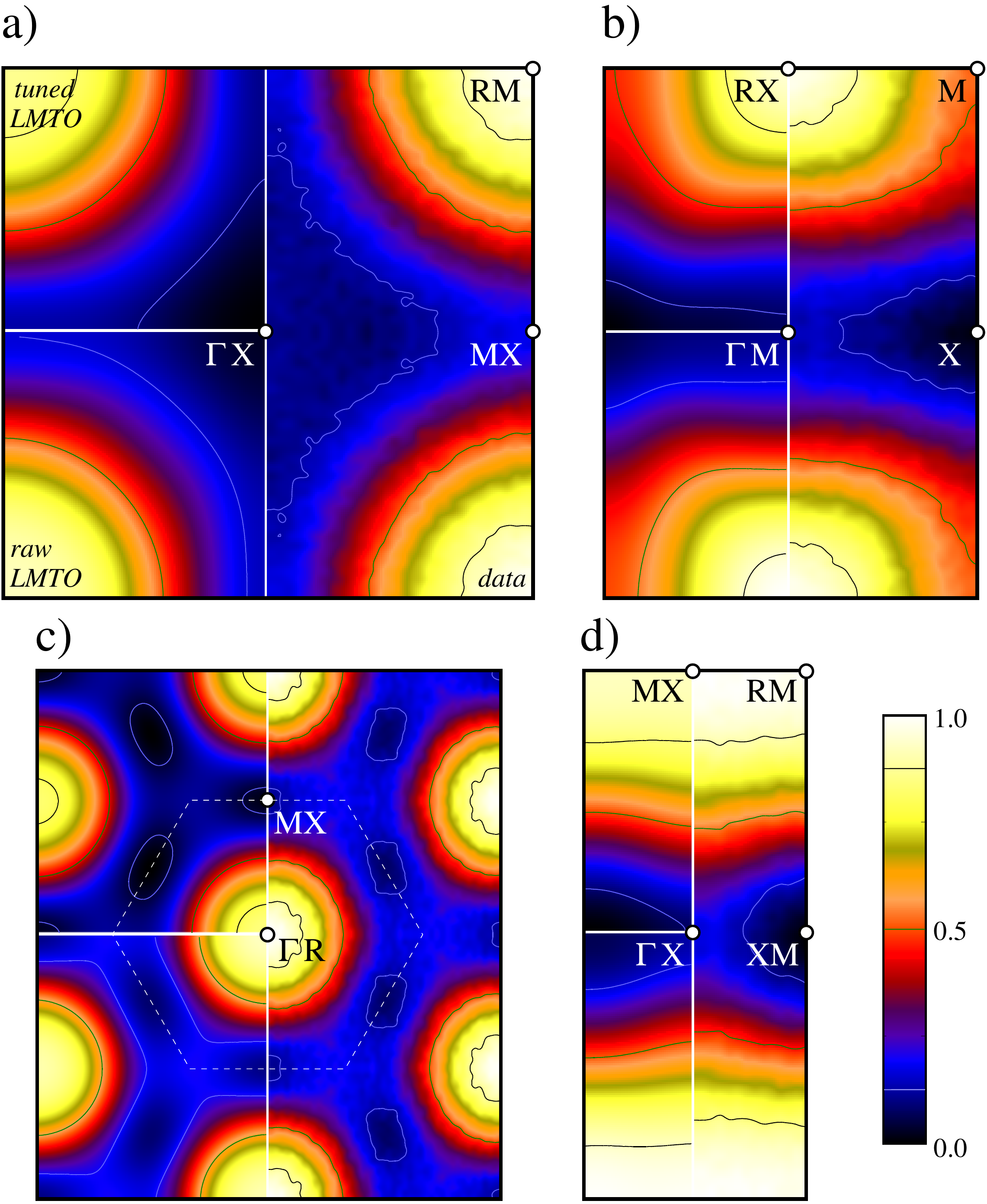}
\end{center}
\vspace*{-0.2in}
\caption{Positron annihilation data in the first BZ for
\mbox{Al-9~at.~\%-Li}, shown in
the right panel, projected along the (a) [100], (b) [110], (c)
[111] and (d) [210] crystallographic directions. The high-symmetry points in
projection have been labelled, and the projected BZ is marked
by the dotted line in (c). In the bottom left panel, the raw LMTO
momentum density is shown, and the tuned calculation
is presented in the top left panel.}
\label{f:data}
\end{figure}

A series of 2D-ACAR measurements along four different crystallographic
directions ([100], [110], [111] and [210]) were made on the Bristol
spectrometer (with a resolution function of $0.16 \times 0.19\;(2\pi/a)$ in the
$x$ and $y$ data axes respectively)
and analysed in order to obtain the projected
electron-positron momentum density in the first BZ, shown in Fig.~\ref{f:data}
(in the absence of positron effects, this would simply correspond to the 
occupation density in the first BZ). The sensitivity of our data to the 
FS is immediately
obvious: Strong regions of high momentum (occupation) density are observed near
the projected $R$-points of the BZ, at which each sheet of FS is fully occupied.
To accompany these measurements, the 3D electron-positron momentum density was
computed from our electronic structure calculations
and the projected (2D) density in the first BZ was obtained
for each corresponding projection.

\begin{figure}[t!] 
\begin{center}
\includegraphics[width=0.84\linewidth,clip]{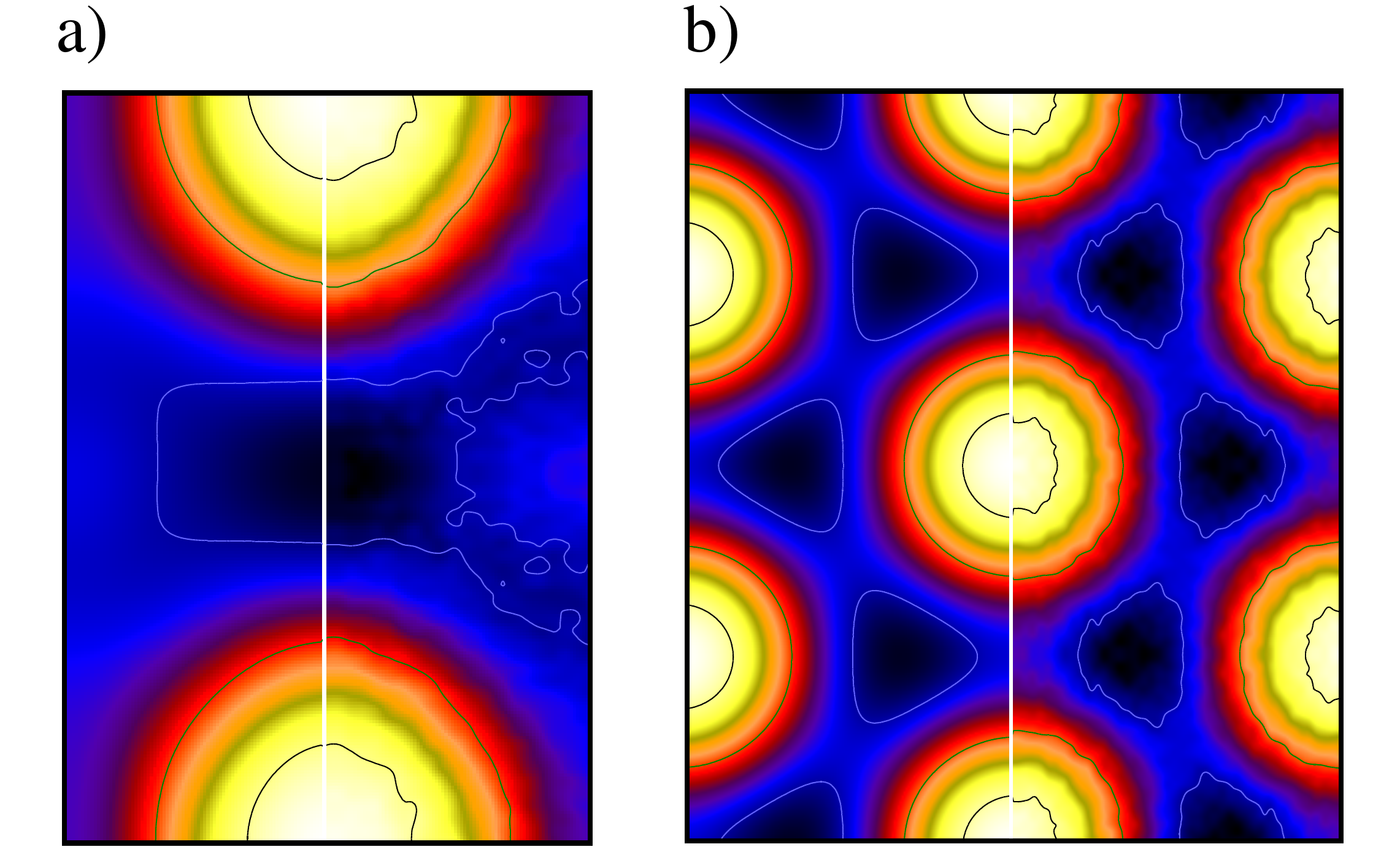}
\end{center}
\vspace*{-0.2in}
\caption{LMTO (left panels) and 2D-ACAR data (right panels) for
pure Al, shown in the $L1_2$ BZ projected along the (a) [110]
and (b) [111] crystallographic directions.}
\label{f:al}
\end{figure}

We begin with a visual comparison of the measured quantities and their
corresponding theoretical distributions. Good agreement is already observed
between the data and the IPM LMTO calculations of Al$_3$Li shown in
Fig.~\ref{f:data}.  However, in order to eliminate the possibility that the
positron
is sampling the Al matrix rather than the $\delta'$ precipitates (scenario ii),
both experimental and theoretical distributions were also obtained of pure Al
along two crystallographic projections ([110] and [111]), shown in
Fig.~\ref{f:al} in the same $L1_2$ BZ of Al$_3$Li. Unsurprisingly, good
agreement is
observed between experiment and theory for Al, but there are stark differences
between the Al distributions and our experimental data for
Al\nobreakdash-9~at.~\%\nobreakdash-Li. In the [110] projection (comparing
Figs.~\ref{f:data}b and \ref{f:al}a), the most obvious differences are in the
shape of the projected FS near $RX$, as well as the distribution near
${\Gamma}M$ and $X$.  For the [111] projection (comparing Figs.~\ref{f:data}c
and \ref{f:al}b), the shape of the distribution, particularly near $MX$, is
markedly different. These results strongly suggest that the positron is, indeed,
sensitive to the Li-rich regions of the crystal, rather than the Al matrix. In
fact, linear combinations of the distributions of the pure Al and $L1_2$
Al$_3$Li calculations were not found to improve the agreement between experiment
and theory over the Al$_3$Li calculations alone, eliminating scenario iii) and
further suggesting that {\em all} of the positrons annihilated from the Li-rich
precipitates, in agreement with independent positron lifetime measurements on
the same samples \cite{jerzy}.

\begin{figure}[t!] 
\begin{center}
\includegraphics[width=0.84\linewidth,clip]{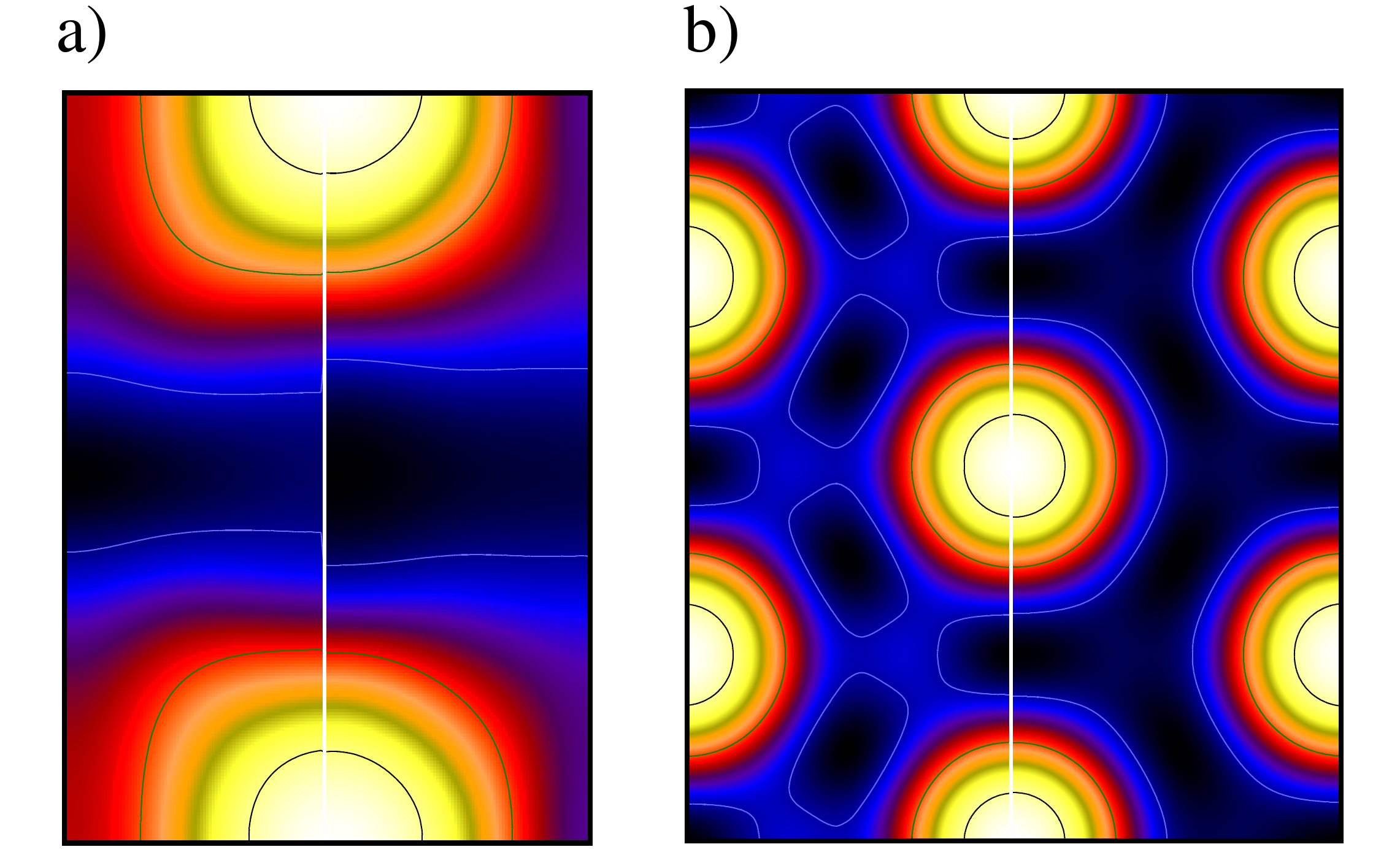}
\end{center}
\vspace*{-0.2in}
\caption{KKR-CPA calculations of the momentum distribution of disordered
Al$_{0.75}$Li$_{0.25}$ (left panels) and Al$_{0.91}$Li$_{0.09}$ (right panels)
projected along the (a) [110] and (b) [111] directions.}
\label{f:kkr}
\end{figure}

In order to address the final alternative possibility (scenario iv), i.e.\ that
the positron is sampling a {\em disordered} alloy of the sample stoichiometry,
KKR-CPA calculations (also employing the IPM) have been performed for disordered
Al$_{0.91}$Li$_{0.09}$, and are shown in Fig.~\ref{f:kkr} for two different
projections (namely [110] and [111]) in the right panels. For the [110]
direction of our data shown in Fig.~\ref{f:data}b, the lowest density is
observed at $X$, and there is a weak peak at ${\Gamma}M$, a feature that is
well-reproduced by our LMTO calculations. In contrast, for the
Al$_{0.91}$Li$_{0.09}$ calculation (Fig.~\ref{f:kkr}a), the minimum density is
located at ${\Gamma}M$, and the shape of the feature at $RX$ is substantially
different. Further inspection of the [111] projection (comparing
Figs.~\ref{f:data}c and \ref{f:kkr}b) yields similar conclusions, leading us to
eliminate this scenario.  Finally, we can consider the possibility that,
although the measurements are sensitive to the precipitate phase, the
precipitates themselves
may not be ordered. However, comparisons with a KKR-CPA calculation of
disordered Al$_{0.75}$Li$_{0.25}$, shown in the left panels of Fig.~\ref{f:kkr},
although superficially similar, are quantitatively in poorer agreement with the
data than the ordered Al$_3$Li calculations. This is most obvious in the
strength of the peaks near ${\Gamma}M$ for the [110], and
near the corner of the projected BZ boundary for the [111] projection.
Together with the comparisons with the LMTO calculations of Al$_3$Li and the
pure Al measurements, this leads us to our first conclusion: that the positron
is sampling the Li-rich $\delta'$ precipitate phase of ordered $L1_2$
Al$_3$Li.

In terms of the FS topology, the agreement between the data and LMTO Al$_3$Li
calculations, although already very good, can be `tuned' by rigidly shifting the
theoretical bands, culminating in a fitted FS whose overall shape more resembles
the experimental FS. This approach has already been successfully applied in
p-space to both positron \cite{major2004a+major2004b} and Compton scattering
\cite{utfeld2009+utfeld2010} data. Here, we operate in k-space, and employ the
state-dependent enhancement scheme outlined in Ref.\ \cite{laverock2010}. This
method, including the site- and orbital-specific annihilation rates to the
positron enhancement as fitting parameters, has been shown to improve the
agreement between theory and high-precision data on the FS of a variety of
elemental metals, as well as providing a measurement of the positron enhancement
itself \cite{laverock2010}.

The results of the rigid-band fit, as expected, reflect the higher positron
affinity of Li over Al, with the positron preferentially annihilating from
within the Li atomic sphere (34.6\%) rather than the equally-sized Al atomic
sphere (21.8\% $\times$ 3). Indeed, this site-dependence is already
well-accounted for by the LDA calculation of the positronic state, which
predicts 35.2\% and 21.6\% respectively for the Li and Al sites. The
state-dependent enhancement factors (see Ref.~\cite{laverock2010}) demonstrate
the de-enhancement of $p$ states (by a factor 0.59) and $d$ states (by a factor
0.43) relative to the $s$ states that has previously been observed for
transition metals and semiconductors \cite{jarlborg1987,barbiellini1997}.

\begin{figure}[t!] 
\begin{center}
\includegraphics[width=1.0\linewidth]{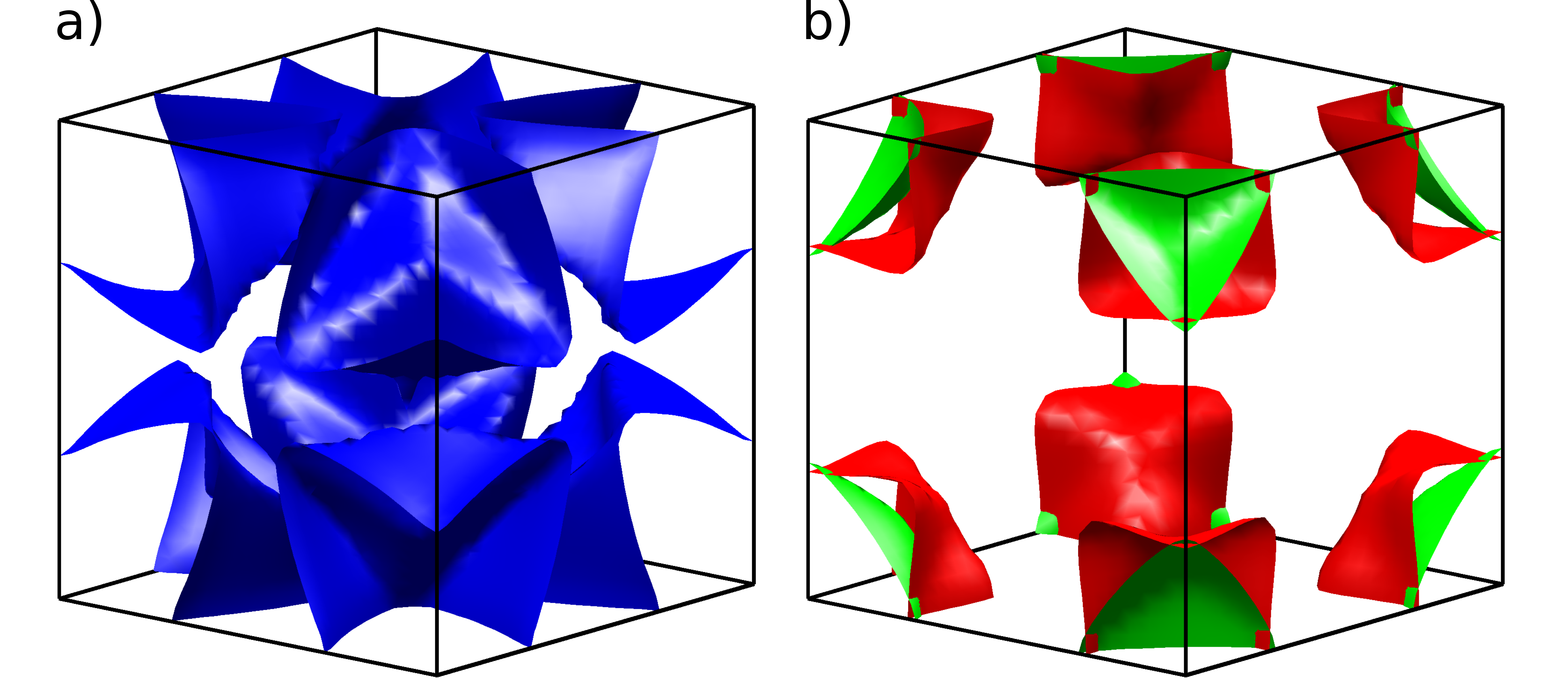}
\end{center}
\vspace*{-0.25in}
\caption{The FS obtained by tuning our LMTO calculation (by rigidly shifting
each band) of Al$_3$Li to the experimental data (compare with
Fig.~\ref{f:fs}).}
\label{f:fsfit}
\end{figure}

Firstly, the rigid-band fit is found to be
sensitive to all of the FS sheets, unambiguously confirming their presence.  The
bands are found to shift upwards in energy (becoming less occupied) by $\sim 50$
mRy, and lead to a `tuned' FS, shown in Fig.~\ref{f:fsfit}. It is emphasized
that there are no constraints to these parameters, meaning that each sheet of FS
is individually free to be completely empty or fully occupied, or indeed any
shape in between allowed by its band structure.
Although the shifts are quite large, the bands are of predominantly $p$
character and have large Fermi velocities (energy gradients), and so
the modification (in k-space) to the Fermi breaks themselves is weaker.
Indeed, we find that this shift in the Fermi wave vector is $\Delta k_{\rm
F} \sim 0.05-0.08\;(2\pi/a)$, which is less than half of the experimental
resolution function.
Nonetheless, these shifts still correspond to a total reduction in the number of
occupied states of $\sim 0.84$~electrons.
The most plausible interpretation of these results is that the topology of
the experimental FS more
closely resembles that shown in Fig.~\ref{f:fsfit} than the raw LMTO
calculation of Fig.~\ref{f:fs}. In previous studies, the present fitting method
has been found to improve the description of the FS (in comparison with
high-precision quantum oscillation measurements) in all the systems investigated
(including Al) \cite{laverock2010}. Moreover, the LDA is well-known to place the
$d$ bands (unoccupied in Al$_3$Li) too low with respect to $sp$ bands
in transition metals \cite{wakoh1973+eckardt1984}.
Our raw LMTO
calculations of Al$_3$Li predict an appreciable hybridization with these
unoccupied $d$ states, having 26~\% total $d$ character at $E_{\rm F}$. In
the rigid-band fit, however, this quantity is somewhat reduced, as the $sp$
bands below $E_{\rm F}$ and $d$ bands above $E_{\rm F}$ are pushed apart. The
results of the fit can be interpreted as a reaction to the over-estimation by
the LDA of the hybridization between the $sp$ and $d$ states, where the flatter
$d$ bands, over-estimated at $E_{\rm F}$, impact on the topology of the FS.
Since the computation of the elastic constants can only be performed at the
minimum of the LDA, the impact these results might have on the predicted
material properties of Al$_3$Li (such as its Young's modulus) remains open to
investigation.

In summary, we have successfully measured the FS of a nano-sized metastable
phase of matter, previously inaccessible using conventional techniques. By
taking advantage of the positron affinity of Li, our 2D-ACAR measurements of
Al\nobreakdash-9~at.~\%\nobreakdash-Li yield a momentum density in close
agreement with electronic structure calculations of ordered $L1_2$ Al$_3$Li,
corresponding to the $\delta'$ precipitates. Moreover, comparisons
with theoretical distributions of pure Al as well as of the disordered alloys
Al$_{0.91}$Li$_{0.09}$ and Al$_{0.75}$Li$_{0.25}$ eliminate other plausible
fates of the positron, firmly establishing the positron as a unique probe of the
$\delta'$ precipitates. Detailed subsequent analysis of the momentum density,
including the application of a novel rigid-band fitting method that additionally
accounts for the state-dependent positron enhancement, has yielded a `tuned' FS
for this elusive phase.

\end{document}